# Predictions of Steady and Unsteady Flows using Machine-learned Surrogate Models

Shanti Bhushan
Dept. of Mechanical Engineering
Mississippi State University
Starkville, MS, USA
bhushan@me.msstate.edu

Greg W. Burgreen
Center for Advanced Vehicular Systems
Mississippi State University
Starkville, MS, USA
greg.burgreen@msstate.edu

Joshua L. Bowman
Dept. Aerospace Engineering
Mississippi State University
Starkville, MS, USA
jb1060@msstate.edu

Ian D. Dettwiller
Engineer Research and Development Center (ERDC)
Vicksburg, MS, USA
ian.d.dettwiller@usace.army.mil

Wesley Brewer
DoD HPCMP PET / GDIT
Vicksburg, MS, USA
Wesley.Brewer@GDIT.com

*Abstract*— **The applicability of computational fluid dynamics (CFD) based design tools depend on the accuracy and complexity of the physical models, for example turbulence models, which remains an unsolved problem in physics, and rotor models that dictates the computational cost of rotorcraft and wind/hydro turbine farm simulations. The research focuses on investigation of the ability of neural networks to learn correlation between desired modeling variables and flow parameters, thereby providing surrogate models. For the turbulence modeling, the machine learned turbulence model is developed for unsteady boundary layer flow, and the predictions are validated against DNS data and compared with one-equation unsteady Reynolds Averaged Navier-Stokes (URANS) predictions. The machine-learned model performs much better than the URANS model due to its ability to incorporate the non-linear correlation between turbulent stresses and rate-of-strain. The development of the surrogate rotor model builds on the hypothesis that if a model can mimic the axial and tangential momentum deficit generated by a blade resolved model, then it should produce a qualitatively and quantitatively similar wake recovery. An initial validation of the hypothesis was performed, which showed encouraging results.**

## I. INTRODUCTION

Computational Fluid Dynamics (CFD) is being widely used as an engineering design tool, as they are inexpensive compared to experiments and allow flow conditions that cannot be replicated in a laboratory. While the advances in high performance computing is pushing the limits of high-fidelity computations (using both large grids and advanced numerical methods and models) for flow physics analysis; the applicability of CFD tools for engineering applications is limited by slow turn-around time. There is a need for low-fidelity models that can be used on coarse grids and provide accurate predictions of integral quantities (such as forces and moments) and large-scale flow features (such as flow separation and progression of vortical structures). In the context of turbulent flows, which involves a wide range of length and time scales and remains a challenge for modelers, the development of low-fidelity models have stalled, and it seems highly improbable that low-fidelity models can be developed without introducing additional complexity (or computational cost). Machine learning (ML) is emerging as a powerful tool that can provide an alternative path to the traditional modeling approach, wherein neural networks can help identify the correlation between input and output flow variables [1].

*A. Background: Machine Learning for Turbulence Modeling*

Availability of numerous direct numerical (DNS)/large eddy simulation (LES) results and experimental datasets is fueling emergence of machine learning as a tool to advance turbulence modeling [2]. For turbulence modeling, the primary unknown is how turbulent stresses are correlated with invariants of mean rate-of-strain tensor and other local flow parameters. The machine learning approach uses a solution database to generate a response surface of desired turbulent feature as a function of input features. A review of the literature shows that the machine learning tools have been used for turbulence modeling for: (1) direct field estimation, wherein the entire flow field is predicted [3,4]; (2) modeling uncertainty estimation, wherein model coefficients are calibrated to minimize modeling errors [5,6]; (3) turbulence model augmentation [7-10]. Turbulence model augmentation is the most common approach used in the literature, and has been applied for unsteady Reynolds Averaged Navier-Stokes (URANS) models either to adjust turbulence production or adjust turbulent stresses or introduce nonlinear stress components; and (4) to obtain a standalone turbulence model, wherein the desired turbulent features are the turbulent stresses. Limited effort has been made in this category [11].

*B. Background: Challenges for Rotor Modeling*

Rotors are used in many engineering applications, such as vertical lift vehicles (family of military helicopters developed by US Army [12]), wind and hydro turbines to name a few. The prediction of the wake of a rotor is critical for engineering design, e.g., for vertical lift vehicles it can affect the loading on the fuselage during maneuvering and can adversely affect its stability; for wind or hydro turbines it dictates the wake recovery length and eventually the array efficiency. CFD simulations with fully resolved blades are the most accurate model for the rotor wake predictions. However, such simulations are numerically expensive, which restricts their applicability during the design cycle. Further, they can be prohibitively expensive for multiple rotor configurations. Several low-fidelity models have been developed to make these simulations more computationally tractable, such as actuator disk or actuator line or actuator surface. Among them, actuator line (ALM) is the most used model, since it provides the best qualitative description of rotor flow. In this model, the blade is modeled as a simple line, and each point on that line imparts a momentum sink corresponding to the lift and drag expected from the blade profile at that location for the local inflow condition, and line rotates similar to the blade [13]. In the general model, the momentum sink is prescribed using a pre-determined look-up table of lift and drag coefficients with respect to angle of attack. These models have two primary limitations: (i) an inability to account for the three dimensionality of the flow or the unsteadiness of the upstream flow; and (ii) an inability to account for turbulence generated by the blade, which is critical for wake growth and recovery. Thus, there is a need for an efficient rotor model that can model both the unsteady momentum sink and turbulence generation from the blades. This research builds on the hypothesis that if a turbine model can mimic the axial and tangential momentum deficit and turbulence generated by a blade resolved model, then it should produce a qualitatively and quantitatively similar mean wake recovery. This research seeks to develop a regression map for the rotor axial and tangential momentum deficit as a first step towards the validation of the hypothesis.

*C. Objectives and Approach*

The objective of this research is to investigate the ability of neural networks to learn a correlation between desired modeling variables and flow parameters, thereby providing a surrogate model that can used in CFD simulations. This investigation is performed for two different classes of problems: turbulence modeling and rotor modeling. For the former, machine learning is used to infer the correlation between the turbulent stresses and invariant of mean rate-of-strain tensor. For the latter, machine learning is used to infer lift and drag expected from the rotor blade profile at any radial location as a function of local inflow condition, such that they can used within actuator-line modeling framework to develop a mid-fidelity rotor model, which can account for inflow unsteadiness.

This research builds on the authors' previous research [1] focusing on development of stand-alone machine-learned (ML) turbulence model for steady boundary layer flow. The key results from the previous research are summarized in the following section. Herein, the above research is extended for unsteady flows. For this purpose, a DNS dataset is obtained for oscillating channel flow, as summarized in Section III.A. In section III.B, the DNS dataset is used for development and validation of a machine-learned turbulence model. In section III.C, data reduction techniques are investigated to minimize skewness of model towards specific turbulence feature due to dominance of that feature in the training dataset. The developed model is validated for both *a priori* and *a posteriori* tests in section III.D. In the *a priori* test, simulation is performed using URANS model, and the ML model is queried as a post-processing step. For the *a posteriori* test, the ML model is coupled with the solver and used for the turbulence modeling. The ongoing work towards the development of the machine-learned rotor model is discussed in section IV. Section V provides some preliminary conclusions and lists future work.

II. PREVIOUS TURBULENCE MODELING RESEARCH

Precursor study [1] focused on development and validation of standalone ML turbulence models for steady boundary layer flow, including assessment of key issues associated with neural network training. To achieve the objectives, two different models were developed: (1) *data-driven* and (2) *physics informed machine-learned* (PIML) [14,15]. The models were applied for the solution of the boundary layer equations:

$$\nu \frac{\partial^2 U}{\partial y^2} + \frac{\partial \tau_{uv}}{\partial y} = u_\tau^2 \qquad (1)$$

where, $\nu$ is the kinematic molecular viscosity, $u_\tau$ is the friction velocity, $U$ is the mean flow velocity and $y$ is the wall-normal direction. The unknown in the above equation is the shear stress $\tau_{uv}$. Physics-based turbulence models express the stress in terms of known flow variables. For example, in the one-equation URANS model [16]:

$$\tau_{uv} = \nu_T \frac{\partial U}{\partial y} \qquad (2)$$

where the turbulent eddy viscosity $\nu_T$ is obtained as:

$$\nu_T = \sqrt{0.3}\ell\sqrt{k} \qquad (3)$$

The above formulation introduces another unknown, turbulent kinetic energy $k$ and $\ell$ is the turbulent length-scale which is expressed in terms of the distance from the wall ($\delta$), such that:

$$\ell = \kappa\delta \qquad (4)$$

where, $\kappa = 0.41$ is the von-Karman constant. The known $k$ is obtained from the solution of an additional transport equation:

$$\frac{\partial k}{\partial t} = \underbrace{\nu_T \left(\frac{\partial U}{\partial y}\right)^2}_{\text{Production}} - \underbrace{\frac{0.1643 k^{3/2}}{\ell}}_{\text{Dissipation}} + \underbrace{\frac{\partial}{\partial y}\left\{(\nu + \nu_T)\frac{\partial k}{\partial y}\right\}}_{\text{Diffusion}} \qquad (5)$$

The ML model seeks to determine a regression map of the turbulent stress in terms of the flow variables, i.e.,

$$\tau_{uv} = f(U, \delta, \nu, u_\tau, \frac{\partial U}{\partial y}) \qquad (6)$$

In the *data-driven* model the regression map for the shear stress was trained to minimize the cost function estimated to be $L_2$ norm error between the machine-learned shear stress and the training dataset. Whereas for the *PIML* model, the cost function included both the errors in the governing equations as well as the $L_2$ norm error.

The study focused on assessment of the following issues:

(a). How to properly train the machine-learned regression map for turbulent stresses? In particular, the effect of input features and training approach on the accuracy of the ML regression map was investigated.

(b). How well does the machine-learned turbulence model satisfy smoothness of the solution during *a posteriori* application, and can they achieve grid independent solution similar to physics-based models?

(c). How does machine-learned turbulence model converge for ill-posed initial condition?

The machine-learned turbulence model was developed from a DNS/LES database of boundary layer flows available in the literature, which included: 21 channel flow DNS cases with Reynolds numbers ranging from $Re_\tau$=109 to 5200 [17]; 18 flat-plate boundary layer with zero-pressure gradient DNS cases with $Re_\theta$= 670 to 6500; and 14 flat-plate boundary layer with zero-pressure gradient LES cases with $Re_\theta$= 670 to 11000 [18]. The database contained around 20,000 data points of which 3.4% were in the sub-layer, 5.7% in buffer layer and 90.9% in the log-layer or the outer layer.

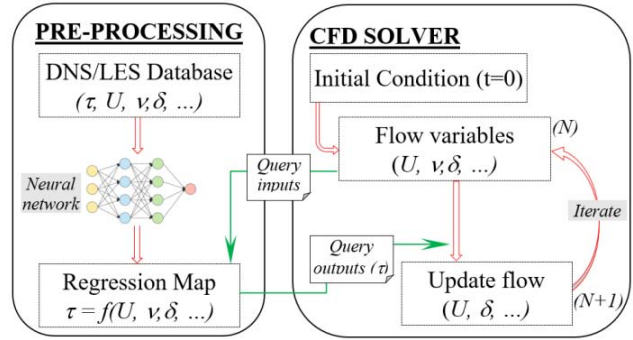

Fig. 1. Schematic diagram demonstrating the coupling of a machine learned turbulence model with CFD solver.

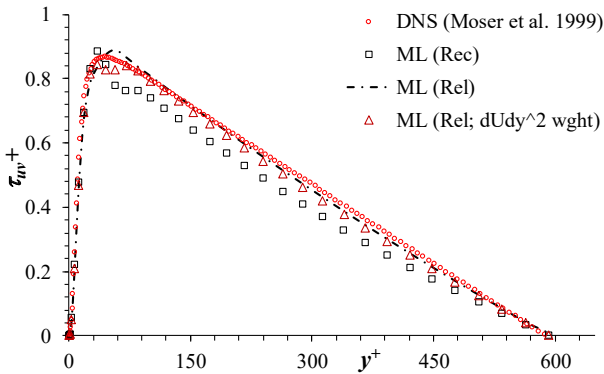

Fig 2. Predictions of turbulent shear stresses for $Re_\tau$ = 590 obtained using *data-driven* turbulence models, where model (*Rec*) is trained using global Reynolds number, (*Rel*) is trained using local flow Reynolds number, and for (*Rel;dUdy^2 wght*), $\left(\frac{dU}{dy}\right)^2$ is used as weights during training.

The machine-learned model was obtained as a pre-processing step using an 8-layer deep neural network with each hidden layer having 20 neurons and a hyperbolic tangent activation function. The model was optimized using an L-BFGS quasi-Newton full-batch gradient-based optimization method and iterated for 2400 epochs.

The models were validated for both *a priori* and *a posteriori* test. In the *a priori* tests DNS results were used as a query inputs, and the query outputs (shear stress) were compared with DNS values. For *a posteriori* tests, the machine-learned turbulent shear stress was used in the solution of boundary layer equation (Eq. 2). Figure 1 provides a schematic diagram demonstrating the coupling of machine-learned turbulence model with CFD solver.

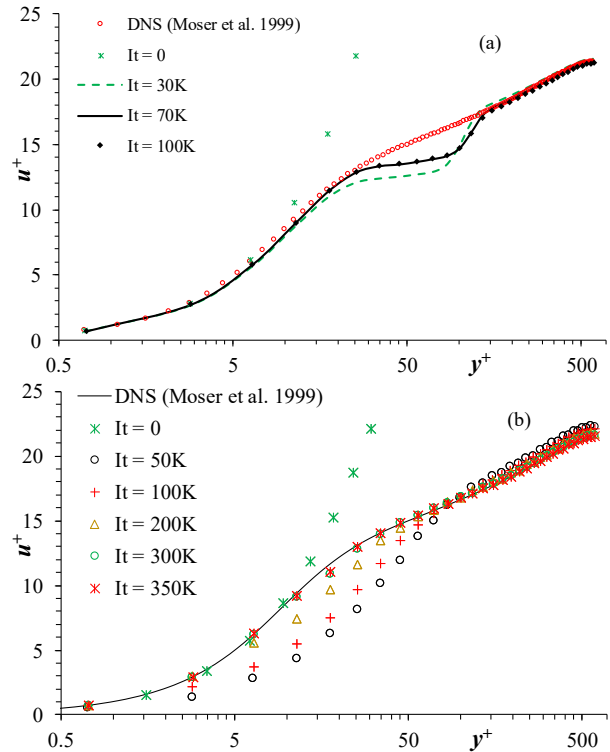

Fig. 3. Predictions of mean velocity for $Re_\tau$ = 590 obtained using: (a) data-driven and (b) *PIML* turbulence models for simulations using $U = (1 - y^8)$ ill-posed initial condition.

The validation study demonstrated that:

(a). The accuracy of the machine-learned response surface depends on the choice of input parameters. Feature engineering was used to find the optimal input features for the neural network training. It was identified that grouping flow variables into a problem relevant parameter improves the accuracy of the model. For example, as shown in Fig. 2, *data-driven* model trained using $Re$ based on local flow velocity and wall distance ($Re_l$) results in 3% error in the shear stress predictions, compared to 12% error for the model trained using $Re$ based on global flow ($Re_c$).

(b). The accuracy of the response surface depends upon how the database is weighted to minimize the overlap between the datasets. This requires a trial-and-error method to come up with an appropriate weighting function. A better way to improve the accuracy of the regression surface is to include physical constraints to the loss function as per [3], i.e., *PIML* approach.

(c). The *data-driven* model showed oscillations in *a posteriori* test, as the query outputs jumped between the database curves. The oscillations were reduced by locally averaging the query inputs and outputs. The *PIML* model did not show such limitations. Since the model query outputs are independent of contiguous data points, thus it was concluded that the smoothness of the solutions depends on the accuracy of the model.

(d). The *data-driven* model predictions converged to unphysical results in the buffer- and lower log-layer when the simulations were started from ill-posed initial conditions, i.e., $U = (1 - y^8)$, where y = ±1 are no-slip walls, as shown in Fig. 3. Note that the flow features in the above profile are outside the training database and the query is essentially extrapolation. Surprisingly, the *PIML* approach worked very well in the extrapolation mode, and the results slowly converged to the correct profile.

### III. OSCILLATING CHANNEL FLOW

Oscillating channel flow is a canonical test case used in the literature to validate the predictive capability of LES models. In this test case an unsteady pressure pulse (body force term) is applied, which introduces periodic unsteadiness in the flow. Scotti and Piomelli [19] performed DNS and LES of pulsating channel flow to study the effect of pressure gradients on the modulation of the viscous sublayer, turbulent stresses and the topology of the coherent structures.

#### A. Direct Numerical Simulations

The DNS simulations in this research are performed using an in-house parallel pseudo-spectral solver, which discretizes the incompressible Navier-Stokes equations using Fast Fourier Transform (FFT) along the homogenous streamwise and spanwise directions and Chebyshev polynomials in the wall-normal direction. The solver is parallelized using hybrid OpenMP/MPI approach, scales up to 16K processors on up to 1 billion grid points, and has been extensively validated for LES and DNS studies [20,21].

DNS were performed for three different (high, medium and low) streamwise pressure pulse, as used in [19]. The simulations were performed using as domain size $3\pi \times 2 \times \pi$ along the streamwise, wall normal and spanwise directions, respectively, on a grid consisting of 192×129×192 points. The domain size is consistent with those used in [19], but grids are finer. The details of the simulation set-up are provided in Table I. The simulations were performed using periodic boundary condition along streamwise and spanwise directions, and no-slip wall at $y = \pm 1$.

The variation of the wall shear stress and streamwise velocity for the medium frequency case is shown in Fig. 4. The positive pressure gradient generates adverse flow conditions and decelerates the flow (and decreases wall shear stress magnitude) and the negative pressure gradient generates favorable flow conditions which accelerates the flow (and increases wall shear stress). The peak wall shear (and velocity) is observed around $3/4^{th}$ cycle. The DNS results were validated in a previous study [22] for the prediction of the alternating (AC) and mean (DC) components of the mean flow against results presented in [19]. The AC components were obtained by decomposing normalized-planar-averaged velocity profile at every one-eight cycle using Fast Fourier Transform at each wall normal location. The DNS results were found to be in good agreement with the available benchmark results.

TABLE I. DNS SIMULATION PARAMETERS. THE NON-DIMENSINAL QUANTITIES ARE HIGHLIGHTED IN YELLOW

| Flow parameters | High frequency | Med. frequency | Low frequency |
|---|---|---|---|
| $Re_{\tau,0}$ | | 350 | |
| $Re_{c,0}$ | | 7250 | |
| $U_c$ | | 1 | |
| $H$ | | 1 | |
| $\nu$ | | 1.38×10$^{-4}$ | |
| Domain size | | $3\pi \times 2 \times \pi$ | |
| Grid | | 192×129×192 | |
| $\partial P_0/\partial x$ | | $u_\tau^2 = 0.002331$ | |
| $\rho$ | | 1 | |
| $u_\tau$ | | 0.048276 | |
| $\alpha$ | 200 | 50 | 8 |
| $\alpha\, dP_0/dx$ | 0.4662 | 0.11655 | 0.01865 |
| $\omega^+ = \omega\nu/u_\tau^2$ | 0.04 | 0.01 | 0.0016 |
| $\omega$ | 0.67565 | 0.16891 | 0.02703 |
| Boundary layer thickness $l_s = \sqrt{2\nu/\omega}$ | 0.2021 | 0.4042 | 1.0106 |
| $l_s^+ = \sqrt{2u_\tau^2/\nu\omega} = \sqrt{2/\omega^+}$ | 7.071 | 14.142 | 35.355 |
| $Re_s = U_o\sqrt{2/\omega\nu}$ | 100 | 200 | 500 |
| $Re_s/l_s^+ = U_o/u_\tau$ | | $10\sqrt{2}$ | |
| $U_o/U_c$ | | 0.03296 | |
| Time step size (Δt) | 0.0002325 | 0.00093 | 0.000969 |
| Timesteps per period (2π/ωΔt) | 40000 | 40,000 | 240,000 |
| Pressure pulse | | $\dfrac{dP}{dx} = \dfrac{dP_0}{dx}[1 + \alpha\cos(\omega t)]$ | |

The variation of the wall shear stress and streamwise velocity for the medium frequency case is shown in Fig. 4. The positive pressure gradient generates adverse flow conditions and decelerates the flow (and decreases wall shear stress magnitude) and the negative pressure gradient generates favorable flow conditions which accelerates the flow (and increases wall shear stress). The peak wall shear (and velocity) is observed around $3/4^{th}$ cycle. The DNS results were validated in a previous study [22] for the prediction of the alternating (AC) and mean (DC) components of the mean flow against Scotti and Piomelli [19] DNS and LES results. The AC components were obtained via a decomposition of the normalized-planar-averaged velocity at eight equidistant times in one cycle by applying FFT. The DC component was the averaged velocity over the entire cycle. The DNS results were found to be in good agreement with the benchmark results.

The low frequency pulse resulted in re-laminarization and transition behavior, which is a challenging case for URANS predictions. The high frequency case was primarily driven by pressure gradient, and turbulence levels were small. The medium frequency case, provides a compromise between the two extremes and was used in this study. For training of the machine-learned turbulence model, the 3D DNS dataset was processed to obtain an unsteady 1D dataset. Due to the use of periodic boundary condition along the streamwise direction, the simulation domain is expected to move in time [23], and as demonstrated in Fig. 5. Further, the results at any time step represents multiple (every grid point in the streamwise-spanwise plane or 192×192) realizations of the turbulent flow field expected at that oscillation phase. The solutions were averaged over these realizations to obtain mean 1D flow along the wall normal (*y*) direction. The 1D solutions every 100th time step or 1/400th pressure oscillation cycle were used to generate a 2D *y-time* map of the mean flow as shown in Fig. 6. Solutions were collected over three pressure oscillation cycles resulting in 129×1201 (or around 155,000) data points.

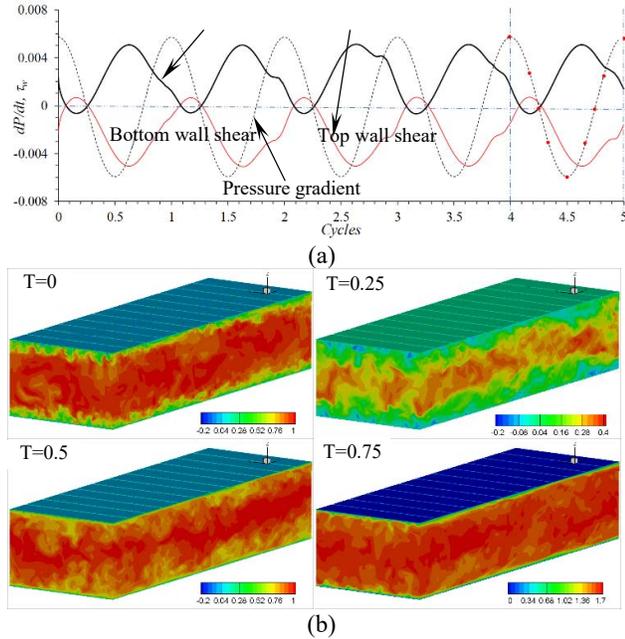

Fig. 4. (a) Variation of wall shear stress over five (5) pressure oscillation cycles, and (b) instantaneous streamwise velocity profile at quarter cycle obtained from DNS for the medium frequency case. T=0 and 0.5 corresponds to the peak and trough of the pressure gradient oscillation.

### B. Machine-learned Turbulence Model

The machine-learned turbulence model is obtained for the mean flow predictions. The governing equation for the mean flow for this case is:

$$\frac{\partial U}{\partial t} = F(t) + \nu \frac{\partial^2 U}{\partial y^2} + \frac{\partial \tau_{uv}}{\partial y} \quad (7a)$$

$$F(t) = \frac{dP_0}{dx}[1 + \alpha \cos(\omega t)] \quad (7b)$$

Similar to the steady state case, the closure of the above equation requires modeling of shear stress $\tau_{uv}$. In addition, for this case, $k$ predictions is also added as a unknown turbulence quantity, and the machine-learned model seeks to obtain regression map with two outputs $(\tau_{uv}, k)$ and the input flow parameters are the local and mean flow quantities as below:

$$\underbrace{(\tau_{uv}, k)}_{Output} = f\underbrace{\left(Re_l = \frac{U\delta}{\nu}; F(t); y^+(t) = \frac{\delta u_\tau(t)}{\nu}; y_0^+ = \frac{\delta u_{\tau 0}}{\nu}; \frac{\partial U}{\partial y}\right)}_{Inputs} \quad (8a)$$

$$u_\tau(t) = \sqrt{\frac{|\tau_w(t)|}{\rho}} \quad (8b)$$

where, $u_\tau$ is the local friction velocity and $u_{\tau 0}$ is the baseline wall friction corresponding to $\frac{dP_0}{dx}$.

For this case only the *data-driven* model was trained, as temporal derivatives could not be computed during training. The model was trained using a 3-layer deep neural network with 512 nodes in each hidden layer with *ReLU* activation functions. The final layer was a linear fully connected layer. The model was trained for 200 epochs using an *ADAM* optimizer with a batch size of 128 and a learning rate of 0.01. During the training, the $L_2$ norm error dropped by an order of magnitude in the first 25 epochs, and the error drop stalled thereafter.

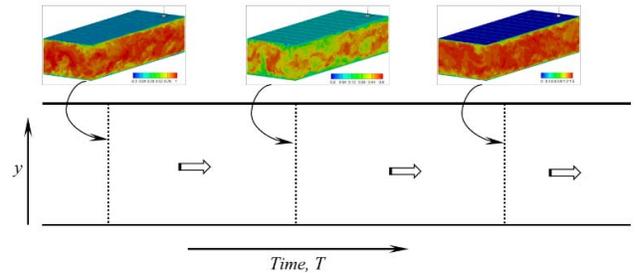

Fig. 5. Schematic diagram demonstrating the post-processing of the 3D DNS data to obtain 1D unsteady mean flow.

### C. Data Clustering and ML Model Refinement

The pulsating channel flow involves a wide range of turbulence regimes unlike the boundary layer case which only has sub-, buffer- and log-layer regimes. As shown in Fig. 6, in this case the turbulence is most prominent towards the end of the pressure-oscillation cycle, and varies significantly along the wall-normal direction. Thus, training a ML model using the entire dataset may be skewed towards the more prominent features. For example, for the boundary layer case the models perform much better in log-layer than in the buffer-layer, as 91% of the datasets were in log-layer region. The *Birch* clustering algorithm *sklearn* was used to cluster the 155,000 data points into unique flow regimes. The clusters were automatically determined using a non-dimensional threshold of 0.3, which resulted in 412 clusters with (min, max, mean, std) of points to be (1, 10700, 376, 937), as shown in Fig. 7. Note that the clustering preserves the periodicity of data.

Several ML models were trained by randomly sampling various percentage of points within each cluster from 0.25% to 100% of the total dataset. As shown in Fig. 8, training using just one datapoint per cluster (0.25% of total) resulted in errors of 35%. The error level reduced to 28% when using four points per cluster (or 1% of total). The error levels were around 12% when 10% of the datasets (either 10 points or 10% of each cluster) was used. A similar error levels where obtained when model was trained using all the datapoints. In addition, training using 10% of the data points required around eight times smaller computational cost compared to those using all of the datapoints. This indicates that an intelligently sampled dataset can generate reliable machine-learned models at a lower computational cost. Note that data clustering can result in significant reduction in computational cost for full four-dimensional (time+ 3D space) datasets.

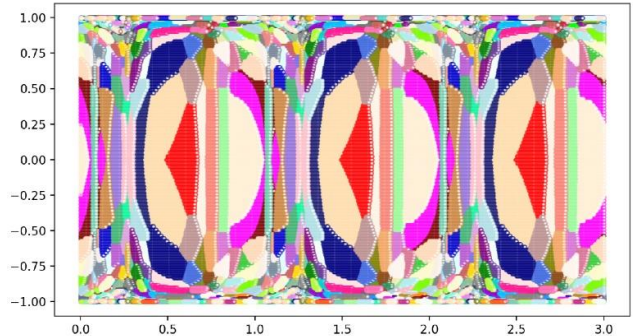

Fig. 7. Clustering of the shear stress DNS data using *sklearn*. The different colored regions represent unique turbulence feature.

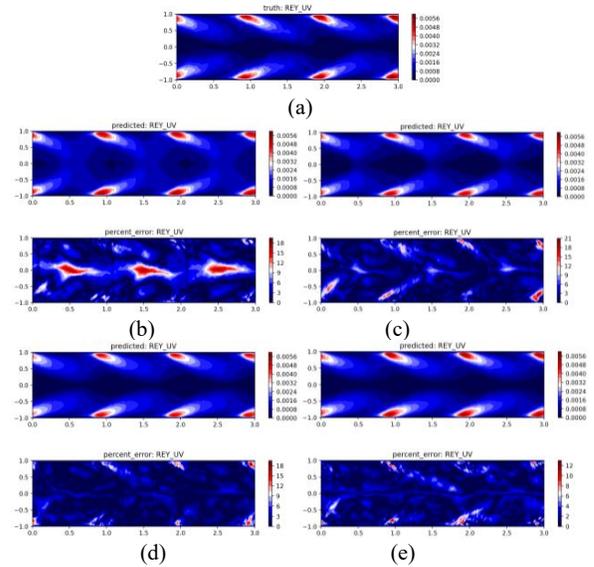

Fig. 8. (a) $\tau_{uv}$ magnitude predicted by DNS. $\tau_{uv}$ regression map (top) and error in the ML model (bottom) trained using (b) 0.25%, (c) 1%, (d) 10% and (e) 100% of datapoints.

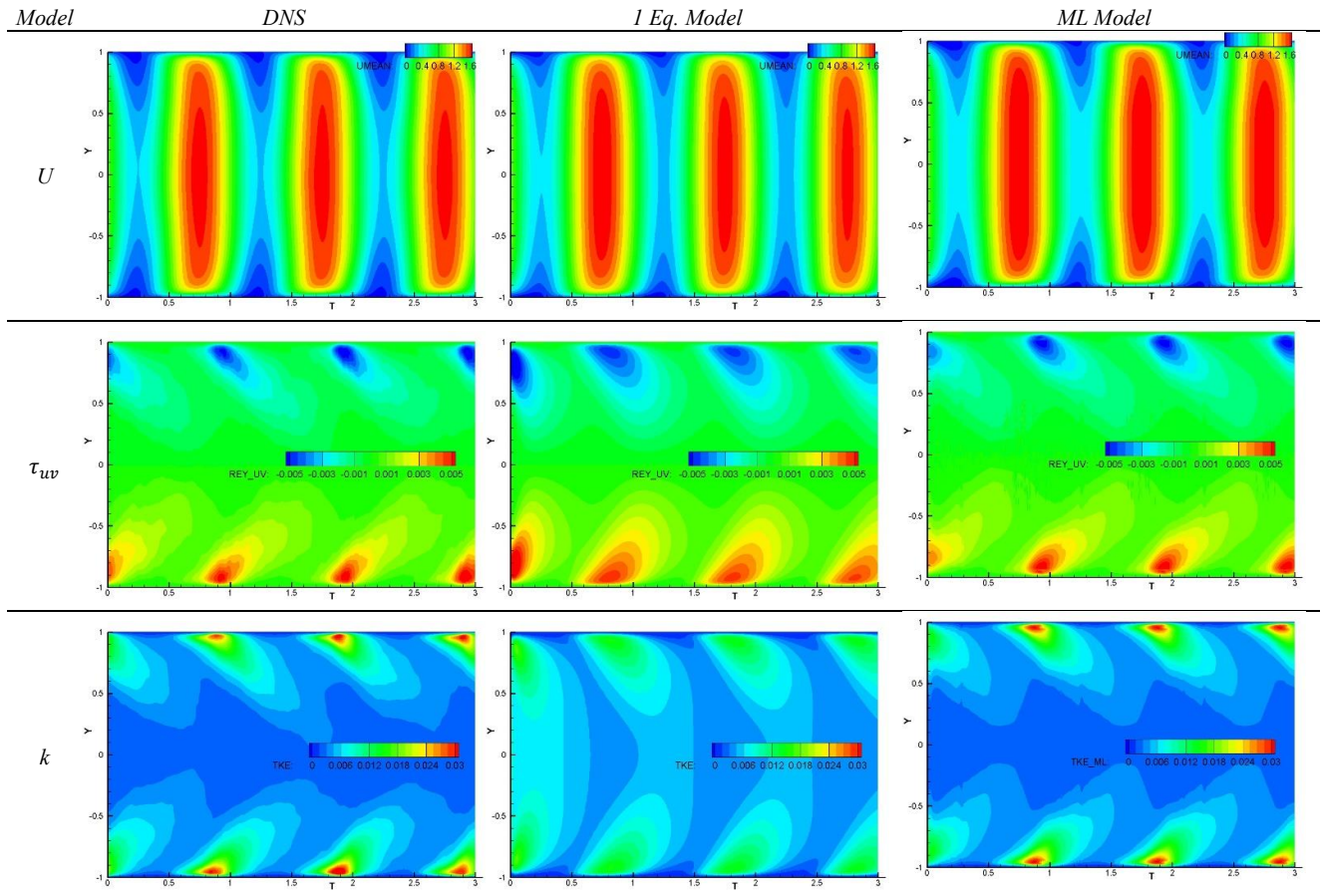

Fig. 6. ML model predictions (right column) are compared with DNS (left column) and URANS (middle column) predictions. The URANS and ML simulations were started from ill-posed initial conditions.

## D. A priori and A posteriori Validation

Three sets of simulations were performed for this case using 1D domain with 65 points in the wall-normal direction. For set #1, the ML model was used in *a priori* mode, i.e., simulations were performed using one-equation URANS model, and the local flow predictions were used to query the regression map. For set #2, the ML model was used in *a posteriori* mode, where the simulations were started from fully converged URANS results at $T=0$. For set #3, the ML was also used in *a posteriori* mode, but the simulations were started from an ill-posed initial condition i.e., $U_{t=0} = (1 - y^8)$. The ML model predictions are compared with URANS predictions and DNS data. Results for simulation set #3 are shown in Fig. 6.

The URANS model performs quite well for the mean flow, but predicts significantly diffused shear stress, and the peak values are 9% underpredicted. The largest error is obtained for the turbulent kinetic energy for which the peak values are underpredicted by 60%. For the simulations started with ill-posed initial conditions, the solutions show large differences with the DNS during the early part of the simulations, but the solution slowly recovers, and the solutions for 2nd and 3rd cycles are very similar to those for the earlier case. This is because the flow is primarily driven by the pressure gradient, and the flow adapts to the pressure variations.

The ML model predictions in *a priori* tests showed significantly better shear stress predictions than the URANS model. The predictions had errors on the order of 3-5% for the mean velocity and shear-stress, and peak TKE were overpredicted by 15%. Since the ML model used the velocities and derivatives predicted by the URANS model, the improved prediction by the former can be attributed to its ability to learn the non-linear correlation between the turbulent stresses and rate-of-strain. The ML model also works very well for the a posterior simulations, and both the mean velocity, shear stress and $k$ compare within 8% of the DNS. The ML model also adjusts to ill-posed initial condition much faster than the URANS model.

## IV. MACHINE-LEARNED ROTOR MODEL

The ongoing research is extending the machine-learned ALM developed for helicopter rotor [24] for solitary hydrokinetic turbine rotor modeling to include the effects of viscous stresses and tip induced turbulence. A feasibility study of the machine-learned actuator line model has been performed, wherein the axial and tangential momentum deficit obtained from the blade resolved simulations are used to specify the lift and drag coefficients of the ALM model.

The preliminary machine-learned ALM (referred to as data-driven ALM) shows very similar instantaneous flow predictions as blade resolved model using just 20% of the computation cost (as shown in Fig. 9). The mean streamwise velocity predictions at the center plane $y/D = 0$ are compared with experimental data [25] and the blade resolved simulation [26] in Fig. 10. The blade resolved model compares reasonably well with the experiment, and predicts the peak deficit behind the blade tips, but the wake diffusion towards the center is somewhat underpredicted. The data-driven ALM model predicts the mean wake deficit very well, and the results are very similar to those predicted by the resolved blade model.

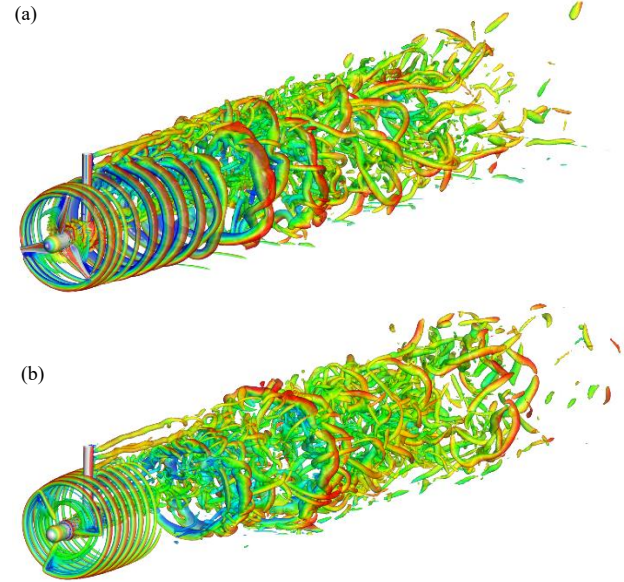

Fig. 9. Prediction of hydrokinetic turbine wake obtained using (a) blade resolved model on a grid with 19M cells, and (b) data-driven ALM on a grid with 12M cells.

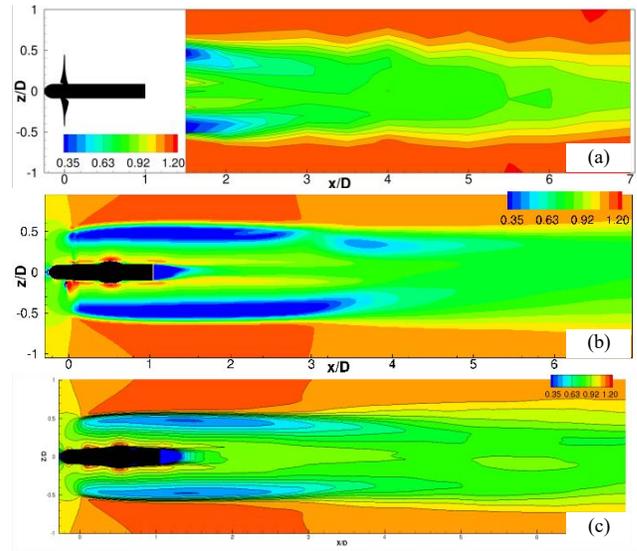

Fig. 10. Mean streamwise velocity contour at centerplane $y/D = 0$: (a) reported in the experiments [25]; (b) blade resolved simulation on a grid consisting of 19M cells [26], and (c) *data-driven* ALM on a grid with 12M cells.

## V. CONCLUSIONS AND FUTURE WORK

The study investigates the ability of neural networks to learn correlation between desired modeling variables and flow parameters for two different class of problems:

turbulence modeling and rotor modeling. For the turbulence modeling, the standalone *data-driven* ML model was developed for unsteady boundary layer flow, which performed significantly better than URANS model and compared within 8% of the DNS. Data clustering is identified to be a useful tool to improve accuracy of the machine-learned model and reduce computational cost. For the rotor modeling, data-driven ALM shows potential in predicting thrust and power, tip vortices and mean wake predictions. Thus, a machine learned ALM as a reliable rotor model is deemed feasible. However, thus far machine learning has been applied for similar types of datasets; which limits the applicability of the model, as well as does not sufficiently challenges the robustness of the machine learning approach.

The ongoing work for ML turbulence model is focusing on generation of a larger database encompassing steady and unsteady boundary layer flows, including separated boundary layer such as flows in converging-divergiong nozzle and over a bump. A model trained using such a database will help in development of a more generic turbulence model for boundary layer flows. It is expected that data clustering will be very helpful for training such as model due to wide range of turbulence characteristics. The ongoing work for the rotor modeling is focusing on generation of database of lift and drag expected from the rotor blade profile for unsteady inflow conditions to train an unsteady ML-ALM.

ACKNOWLEDGMENT

Effort at Mississippi State University was sponsored by the Engineering Research & Development Center under Cooperative Agreement number W912HZ-17-2-0014. The views and conclusions contained herein are those of the authors and should not be interpreted as necessarily representing the official policies or endorsements, either expressed or implied, of the Engineering Research & Development Center or the U.S. Government.

This material is also based upon work supported by, or in part by, the Department of Defense (DoD) High Performance Computing Modernization Program (HPCMP) under User Productivity Enhancement, Technology Transfer, and Training (PET) contract #47QFSA18K0111, TO# ID04180146.